\documentclass[preprint,12pt]{elsarticle}
\usepackage{graphics,bm}
\usepackage{amssymb}
\usepackage{epsfig}
\usepackage{epsf}
\usepackage{graphicx}
\usepackage{mathrsfs}
\usepackage{amsfonts}
\makeatletter

\newcommand{\beqa}{\begin{eqnarray}}
\newcommand{\eeqa}{\end{eqnarray}}
\newcommand{\beq}{\begin{equation}}
\newcommand{\eeq}{\end{equation}}

\newcommand{\nc}{\mathbf{\rm\bf nc}}
\newcommand{\sn}{\mathbf{\rm\bf sn}}

\newcommand{\prb}{Phys. Rev. B}
\newcommand{\pra}{Phys. Rev. A}
\newcommand{\prl}{Phys. Rev. Letters}

\newcommand{\mbf}[1]{\mbox{\boldmath$#1$}}

\journal{Physics Letter}
  
\begin{document}
\begin{frontmatter}

\title{Josephson Effects in a Bose-Einstein Condensate of Magnons}

\author[rvt,focal]{Roberto E. Troncoso}
\ead{r.troncoso.c@gmail.com}

\author[focal]{\'Alvaro S. N\'u\~{n}ez}
\address[rvt]{Centro para el Desarrollo de la Nanociencia y la Nanotecnolog\'ia, CEDENNA, Avda. Ecuador 3493, Santiago 9170124, Chile}
\address[focal]{Departamento de F\'isica, Facultad de Ciencias F\'isicas y 
Matem\'aticas, Universidad de Chile, Casilla 487-3, Santiago, Chile}
\ead{alnunez@dfi.uchile.cl}
\begin{abstract}
A phenomenological theory is developed, that accounts for the collective dynamics of a Bose-Einstein condensate of magnons. In terms of such description we discuss the nature of spontaneous macroscopic interference bet	ween magnon clouds, highlighting the close relation between such effects and the well known Josephson effects. Using those ideas, we present a detailed calculation of the Josephson oscillations between two
magnon clouds, spatially separated in a magnonic Josephson junction.
\end{abstract}


\end{frontmatter}

%++++++++++++++++++++++++++++++++++++++++++++++++++++++++++++
\section{Introduction} 
Efforts to improve our understanding and ability to manipulate magnonic excitations  in ferromagnetic thin films have received great attention in recent years \cite{Kruglyak,Melkov}. A special role in this context is played by Ytrium-Iron garnet (YIG) based systems. Characterized by a magnon spectrum greatly isolated from other degrees of freedom \cite{SAGA,Serga}, YIG systems have allowed the observation of several novel effects (magnonic lattices, spin pumping, etc.).  In this work we focus on several reports on the Bose-Einstein condensation (BEC) of magnons \cite{Demokritov2006, Demidov1, Demidov2}.

Along with the uncontroversial evidence of macroscopic occupation of the lowest lying state, several questions arose regarding the appropriateness of the concept of BEC to refer to collective magnon behavior \cite{Snoke}. One of the most striking phenomena in the nature of BECs is the emergence of a macroscopic wavefunction that displays phase coherence over macroscopic length-scales. Finding effects associated with such coherence is important in clarifying the true nature of cloud of condensed magnons. Recently in \cite{Demokritov0} the existence of the two-component Bose-Einstein condensate of magnons was reported as the linear superposition between two spatially non-uniform macroscopic wavefunctions, describing the magnon-BEC state at the doubly degenerate lowest-energy. This interference results in a spatially non-uniform ground state with a periodic modulation at the density profile of the condensate, the so-called {\it spin density wave}(SDW), as correctly predicted early in \cite{Troncoso} and recently studied in \cite{Fli}. Additionally, in \cite{Demokritov0}  vortex-like excitations in the gas of condensed magnons were observed. The latter reflect themselves as dislocations at the SDW pattern. This type of topological excitations was first studied in \cite{Troncoso} based on a microscopic model for the condensed of magnons. 

An interesting discussion was carried out in \cite{Fli} where the authors study the relation between the magnon-BEC phase transition and the contrast of the experimentally observed periodic modulation. In particular, it is suggested that the phase transitions may be identified by measuring the contrast of the spatial interference pattern for different values of the thickness of the sample $d$ and the in-plane magnetic field $H$. Additionally, they addressed a type of collective mode referred as zero sound in analogy to the Landau's Fermi liquid theory. This oscillation results from the coupling of the relative phase between both components of the condensate and its imbalance density.

In the present work we start from a phenomenological stand-point and proceed to explore the physical nature of the non-linear dynamics of the condensate. To exploit the occurrence of the macroscopic coherence is necessary to analyze and perform the macroscopic interference effect between magnon condensates. The interference phenomena of such states is referred as the Josephson effect. This fundamental issue is devoted to explore the superfluid properties of the magnon condensate as well as to establish a way to settle the controversial aspects, which are related to the true nature of the condensate of magnons  \cite{Ruckriegel}, from the experimental point of view. 

Discovered and observed early in superconductivity \cite{Josephson, Shapiro}, the Josephson effect has been observed in superfluid helium ${}^{3}$He \cite{Backhaus} and ${}^{4}$He \cite{Sukhatme}, and in Bose-Einstein condensates of alkali atomic gases in double well traps \cite{Levy}. In the last case, the Josephson dynamics between weakly coupled BEC's manifests itself in several novel phenomena, mainly due to the nonlinearity that stems from the interaction among bosons. The observation of such phenomena in the context of magnon condensate will provide irrefutable evidence of the spontaneous macroscopic coherence.

The realization of the magnonic Josephson junction (MJJ) will consist principally of two stages. The first is based on the usual way for modeling the splitting of condensed clouds in alkali atomic gases, i.e. introducing a potential well inside the trap that splits the single trapped condensate into two parts. The partitioning leads to two weakly coupled condensates, where the tunneling can be tuned modifying the parameters of the system. The second point is related to the spin-wave tunneling effect on ferromagnetic thin films. In that situation a magnetic field inhomogeneity is induced over the thin film by a conductor placed transversely. The magnon condensate created on the film will be divided into two parts when, by the wire conductor, flows a DC current in such direction that increases locally the magnetization. 

Within this work the dynamics of condensed magnons, in a double-well potential, will be described by a Gross-Pitaevskii-like equation \cite{Bose-Einstein-General}. This equation will be derived phenomenologically. It turns out to be essentially the same as the one that can be derived microscopically \cite{Troncoso}.

\section{Phenomenological Description of the Magnon Condensate}\label{sec: phenomenological approach}
\begin{figure}[htbp] %  figure placement: here, top, bottom, or page
   \centering
   \includegraphics[width=0.7\textwidth]{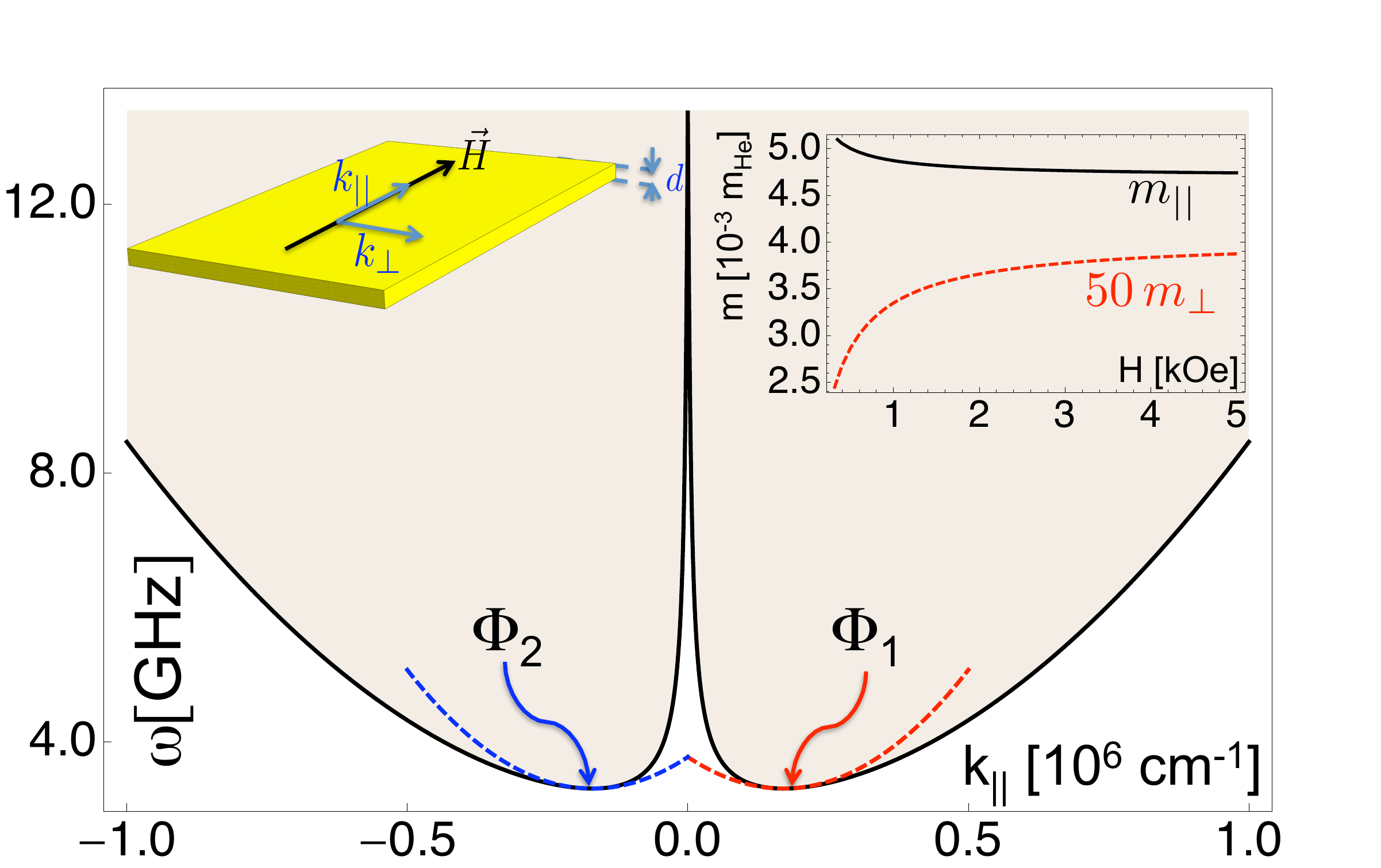} 
   \caption{Spectrum of magnons in a YIG thin film, with an in-plane field of $H\sim 1 kOe$ and film thickness $d\sim 5 \mu m$, for momentum parallel to the field. The continuos line correspond to the spectrum as presented in \cite{Rezende}. Magnons accumulate in the vicinity of the two minimum energy states around which the dispersion is accurately described by a quadratic form depicted by dashed lines. {\em Inset-} Effective masses around the minimum energy states as a function of the external magnetic field. The anisotropy manifest itself in distinct masses for spin waves with momenta along  and perpendicular to the external magnetic field. The masses differ by a factor  of about $10^2$ rendering the magnon system as highly anisotropic.}
   \label{fig:spectrum}
\end{figure}

In a related work \cite{Troncoso} we have discussed theoretically the existence of phase coherence starting from the standard microscopic description of the magnon gas dynamics. Our theoretical description predicts the existence of such coherence when the magnon density reaches a certain critical density. 
Interactions between magnons turned out to be essential in the creation of such coherence. In this work we pursue a phenomenological approach, based on the basic microscopic features that gave rise to our previous treatment. The basic conclusions reached at the end of both treatments are essentially equivalent. The basic processes that have been found \cite{Akhiezer, Kalinikos} to dominate magnon dynamics are: 
\begin{enumerate}[(i)]
\item a dipolar interaction-renormalized dispersion relation that shifts the states of minimum energy away from the ${\bf k}=0$, that is expected solely on 
account of the exchange term, to ${\bf k}=\pm{\bf k}^{0}$. This degenerate minima is depicted in Fig. \ref{fig:spectrum}; 
\item a so-called 3-magnon confluence (resp. splitting) term that 
reduces (resp. increases) the magnon number. These processes are consequence of the long wave length contributions of the dipolar energy;  
\item  a magnon-magnon scattering term that comprises contributions of both the exchange and the dipolar interactions,
\item parametric excitation of magnons, through a pumping field that creates magnons at a rate, $\mathcal{P}$. Magnon condensation, in the form of macroscopic occupation of the lowest energy state, is observed when $\mathcal{P}$ exceeds a critical value, $\mathcal{P}_c$.
\end{enumerate}
We note that magnon excitations can be treated effectively as bosonic excitations. Indeed, we can  use bosonic operators  directly related to the magnetization through the well known Holstein-Primakoff transformation \cite{Holstein,Auerbach}. 
In this representation the spin ladder operators are mapped into bosonic creation and annihilation operators. In this way the spin raising operator  is associated with the annihilation of a bosonic excitation $S^+_i\sim b$, while the spin lowering operator is correspondingly associated with the creation of a bosonic excitation $S^-_i\sim b^\dagger$.
The dispersion relation can be written, in the vicinity of the base states, in terms of effective masses, 
\beq
\hbar\omega_{\bf k}=\hbar\omega_0+\hbar^2{q^2_{||}}/2 m_{||}+\hbar^2{q^2_{\perp}}/2 m_\perp,\label{eq: spectrum}
\eeq
 where ${\bf q}={\bf k}\pm{\bf k}^{0}$. 
We remark the explicit global $U(1)$ symmetry breaking induced  by magnon decay processes on this model. This peculiar  behavior reflects the fact that the full dipolar interaction term does not conserve 
the net magnon number. It will be shown that this fact does not pose any obstacle to a proper interpretation of the system's behavior in terms of spontaneous coherence phenomena. This fact is in direct analogy with the case of the magneto-crystalline anisotropy in ferromagnets. As in that case, the weak
anisotropy is of relevance only after a condensate state is achieved.

To take into account macroscopic coherence over macroscopic length scales, an envelope wavefunction approach can be envisaged. From this picture, the system is described in terms of the two collective wave-functions associated with each minima. Using the collective field $\Phi_\sigma(x,t)$, whose absolute value corresponds to the local density of magnons in states $\sigma=-1,1$, see Fig. \ref{fig:spectrum},  while its phase correspond to the local collective phase. 
The energy associated with this state can be written in a compact form by using the following notation, $(x,t,\sigma)$ labels will be summarized in a single subindex. 
For a homogeneous system the energy can be expressed in terms of the series expansion
\beqa
{\mathcal E}&=&\sum_{M=m+n}\Gamma^{\sigma,x,t}_{\eta,y,\tau}\;\Phi_{\sigma_1}(x_1,t_1)\cdots\Phi_{\sigma_n}(x_n,t)\Phi^*_{\eta_1}(y_1,\tau_1)\cdots\Phi^*_{\eta_m}(y_m,\tau_m)
\eeqa
Terms that are unbalanced in the field and its conjugate are in explicit violation of overall $U(1)$ symmetry associated with conservation of the number of particles. In general, this argument forces them to cancel. However, the microscopic dynamics of magnons does not manifest invariance under such symmetry, reflecting the inherent lack of magnon conservation, and in principle such terms must be considered. By restricting our attention to low momentum, we need to focus only on those terms for which $\left(\sigma_1+\cdots+\sigma_n\right)=\left(\eta_1+\cdots+\eta_m\right)$. In particular, we can discard from the contribution to the energy, terms proportional to odd powers of the fields. Such reduction is as far as one can get due to the $U(1)$ symmetry-breaking terms.
The "anomalous" terms are restricted to valley-mixing terms. Requiring that: (1) in the limit of vanishing density the system recovers the magnon spectrum Eq. (\ref{eq: spectrum}), (2) the net momentum of the magnons is zero, and (3) the system is symmetric with respect to valley indices;  it is possible to simplify the energy into:
\beqa
\mathcal{E}[\Phi,\Phi^*]&=&\int d{\bf r}\left(
\left(
\Phi^{\dagger}_1\;\hbar\omega({\bf \partial_r})\Phi^{\phantom\dagger}_1+
\Phi^{\dagger}_2\;\hbar\omega({\bf \partial_r})\Phi^{\phantom\dagger}_2
\right)+\mu \left(\Phi^{\dagger}_1\Phi^{\phantom\dagger}_1
+\Phi^{\dagger}_2\Phi^{\phantom\dagger}_2\right)\right.\label{eq: eff action}\\
&+&\mbf{\nu}\;\Phi^\dagger_1\Phi^\dagger_2+
{\mbf{\nu}}^{*}\;\Phi^{\phantom\dagger}_1\Phi^{\phantom\dagger}_2+\frac{\gamma_1}{2}\left(\Phi^{\dagger}_1\Phi^{\phantom\dagger}_1+\Phi^{\dagger}_2\Phi^{\phantom\dagger}_2\right)^2+\left.\frac{\gamma_2}{2}\left(\Phi^{\dagger}_1\Phi^{\phantom\dagger}_1-\Phi^{\dagger}_2\Phi^{\phantom\dagger}_2\right)^2\right)\nonumber
\eeqa
where $\mbf{\nu}$, $\gamma_1$ and $\gamma_2$ are phenomenological parameters that should be determined from the experiment. Despite the explicit breakdown of the $U(1)$ symmetry, as reflected by the terms proportional to $\mbf{\nu}$,
this energy is invariant under the residual symmetry transformation:
\beqa
\Phi_1\rightarrow{\rm e}^{i\delta}\Phi_1\;\;\;\; &\mbox{and, }\;\;\;\;&\Phi_2\rightarrow{\rm e}^{-i\delta}\Phi_2.
\eeqa
The parameters $\mu$, $\bar{\nu}$, $\gamma_1$ can be obtained from experimental data as follows. First we set $\mbf{\nu}=\bar{\nu}{\rm e}^{i\psi_\nu}$ If $\gamma_2>0$ the energy is easily minimized by equally populating both valleys. 
Let $\Phi_\sigma=\sqrt{n}{\rm e}^{i\psi_\sigma}$, the energy density becomes
\beq
\frac{\mathcal{E}}{A}=2n \left(\mu+\bar{\nu}\cos\left(\psi_1+\psi_2+\psi_\nu\right)\right)+2\gamma_1 n^2
\eeq
From the last equation, we find a condensation transition at $\mu-\bar{\nu}<0$. We identify this symmetry breaking transition with the transition towards a macroscopically occupied lowest energy state reported in the experiments \cite{Demokritov2006, Demidov1, Demidov2}. 
We can use this fact to associate 
\beq\label{eq: transition}
\mu-\nu=\lambda\left(\mathcal{P}_c-\mathcal{P}\right),
\eeq 
for a positive value phenomenological parameter $\lambda$ and where ${\cal P}_c$ is the critical pumping power to reach the condensation transition.Based on the Eq. (\ref{eq: transition}) we interpret the anomalous coefficient $\bar{\nu}$ as composed of $\nu$, coming from the intrinsic dynamics of the gas, and ${\cal P}$ the external flow of magnons into the condensate, i.e., the parametric pumping. The stationary density of magnons in such regime is: 
\beq
n_{\rm BEC}=2n=\frac{\lambda}{\gamma_1}\left(\mathcal{P}-\mathcal{P}_c\right).
\eeq 
Remarkably, this linear behavior in $\left(\mathcal{P}-\mathcal{P}_c\right)$ is a natural consequence of the phenomenological approach together with the identity in Eq. (\ref{eq: transition}). Its agreement with experimental data \cite{Demokritov2} can be readily verified and constitutes a non trivial correct prediction of the present phenomenological model.
Additionally, following \cite{Bose-Einstein-General},  a healing length can be calculated:
\beq\label{eq: healinglenght}
\zeta^2=\frac{\hbar^2}{2\sqrt{m_{||}m_\perp}\lambda}\left(\mathcal{P}-\mathcal{P}_c\right)^{-1}.
\eeq
with a measurement of $ \mathcal{P}_c$, $n_{\rm BEC}$ and $\zeta$ the phenomenological parameters can be determined.

The main result of this section is to provide a phenomenological picture of the collective dynamics of the magnons. The dissipation mechanisms can be encoded, within the phenomenological approach, in terms of a Rayleigh dissipation function \cite{Landau}. In principle, this function must be expanded in powers of $\partial_t \Phi_i$, this expansion to lowest order becomes:
\beq
\mathcal{R}=\alpha\int d{\bf r}\left(\left|\partial_t \Phi_1\right|^2+\left|\partial_t \Phi_2\right|^2\right).
\eeq
where $\alpha$ characterizes the damping constant as a phenomenological parameter.

The condensate consists, roughly speaking, of two magnon condensates lying the vicinity of the two points of minimum energy in momentum space, and magnetic interactions introduce a coupling between them. The basic phenomenological description of the dynamics is obtained, using the energy functional $\mathcal{E}$ together with a kinetic term 
\beq
\mathcal{S}=\int d{\bf r}d{t} \left(\Phi^{\dagger}_1 i\hbar\partial_t\Phi^{\phantom\dagger}_1+\Phi^{\dagger}_2 i\hbar\partial_t\Phi^{\phantom\dagger}_2\right)-\mathcal{E}[\Phi,\Phi^*].
\eeq
In the magnon condensate, the equations of motion correspond to the Euler-Lagrange equations:
\beq
\frac{\delta \mathcal{S} }{\delta\Phi^\dagger_i}=\frac{\delta \mathcal{R} }{\delta\left( \partial_t\Phi^\dagger_i\right)}.
\eeq
Straightforward calculations lead to the conclusion that the dynamics of the two condensates can be described by the following generalized pseudo-spin GPE:
\beqa
i\hbar (1+i\alpha)\partial_t|\Psi\rangle&=&-\frac{\hbar^2}{2 m_{||}} \nabla_{||}^2 |\Psi\rangle-\frac{\hbar^2}{2 m_{\perp}} \nabla_{\perp}^2 |\Psi\rangle + \mu |\Psi\rangle\nonumber\\+\;\;  \bar{\nu}\sigma_{x}|\Psi^*\rangle&+&\gamma_1 |\Psi|^2|\Psi\rangle +\gamma_2 \langle\Psi|\sigma_z|\Psi\rangle\sigma_z|\Psi\rangle \label{eq: gross-pitaev}
\eeqa
where the pseudo spin $|\Psi\rangle=(\Phi_1,\Phi_2)^t$ refers to valley degeneracy in momentum space,  while $\alpha$, $m$, $\mu$, $\bar{\nu}$ and $\gamma_i$ are real parameters characterizing  the dynamics. 
The only term in the equation that breaks the time reversal symmetry (associated with the transformation $|\Psi(t)\rangle\rightarrow\sigma_x |\Psi^*(-t)\rangle$) is the term proportional to $\alpha$. This term plays the role of a damping constant much in the same way as the Gilbert damping term in magnetism \cite{Qian,JFR,Rossi}. We note that this equation presents an interesting phenomenology that shares with other novel phenomena, e.g., the spin-orbit coupled Bose-Einstein condensates \cite{sobec}. Before closing this section we comment on other phenomenological approaches that have been taken in the literature. Gross-Pitaevskii equations have been constructed to describe the dynamics of magnon condensates in the works of \cite{Malomed, Rezende2} and recently in \cite{Fli}. We emphasize that this form of the equation is essentially different from the phenomenological ones proposed in those works, since Eq. (\ref{eq: gross-pitaev}) has a different form for the dissipation term and an explicitly gauge symmetry breaking term proportional to $\bar{\nu}$. In the next section a discussion concerning with the experimental relevance of Eq. (\ref{eq: gross-pitaev}) will be carried out.

\section{Stationary states and comparison with experiment}
Several predictions that can be drawn from the phenomenological model just described have been experimentally ratified. We start with the prediction of a spin density wave whose period is determined by the interference between the condensates at different valleys.

The existence of periodic modulation of the condensate density, i.e., {\it spin density wave}, as well as its dislocation (equivalent to the vortices of the condensate) was recently observed in the experimental work \cite{Demokritov0}. This detection clearly shows the phase blocking between the two condensates located at $\pm{\bf k}^{0}$ implied by the breaking of the residual symmetry discussed above. In what follows, we will discuss the experimental results in the light of our phenomenological model. We will show that several features of the experiment can be given a simple explanation within the language presented in the previous section.

As well described in the last section, the magnons spontaneously occupy macroscopically the doubly degenerate lowest energy when the condition Eq. (\ref{eq: transition}) is satisfied. The two-component wave function for the ground state is directly associated in space by the periodic modulation of the magnetization deviation $\delta{\bf M}= \sqrt{\rho}(\cos\theta,\sin\theta,0)\cos\left({k}^{0}x+\delta\right)$. The phase $\theta$ defines the plane of polarization of the SDW, while its position is fixed by the relative phase $\delta$ and, according to the previous discussion, determined by the spontaneously breaking symmetry mechanism. The wavevector of the condensed magnons determines the wavelength of the SDW and this is given by $\lambda_{SDW}=2\pi/|{\bf k}^{0}|$. For the experimental parameters used in \cite{Demokritov0}, i.e., a YIG's thin film of $5.1[\mu m]$ thickness and placed into a static magnetic field $H_0=0.1 T$, we estimate for SDW wavelength $\lambda_{SDW}\approx 1.8[\mu m]$ which is very similar to the experimental result (See Fig. $1$ of Ref. (\cite{Demokritov0}). 

A consequence of long-scale phase coherence is the emergence of vortex-like excitations from the ground state of the condensate, which is a natural feature inherited from the breaking of the $U(1)$ symmetry. This is observed as a dislocation-like defects in the SDW pattern \cite{Demokritov0}. This kind of excitations was studied in \cite{Troncoso} where the vortex is described by a velocity field ${\bf v}=\frac{\hbar}{\sqrt{m_{\perp}m_{||}}}\nabla\delta({\bf x})$, with $\delta({\bf x})$ the phase of the condensed magnons. The single-valuedness of the wavefunction for the vortex, given by $\Phi_{(1,2)}({\bf x})=\sqrt{\rho_0}{\cal R}({\bf x})e^{\pm\ell\delta({\bf x})/2}$, requires that $\ell$ be an integer (where $\ell$ is the winding number) with $\sqrt{\rho_0}$ being the condensate density in the ground state. The profile density ${\cal R}({\bf x})$ vanishes in the origin, since the velocity field is singular at $|{\bf x}|\rightarrow 0$, and becoming constant in the bulk. The size of the vortex is determined by a characteristic length, the so-called healing length. Such description is equivalent to the phenomenology presented in the previous section and it is fully characterized by the Eq. (\ref{eq: healinglenght}). 
\begin{figure}[htbp] 
   \centering
   \includegraphics[width=2.8in]{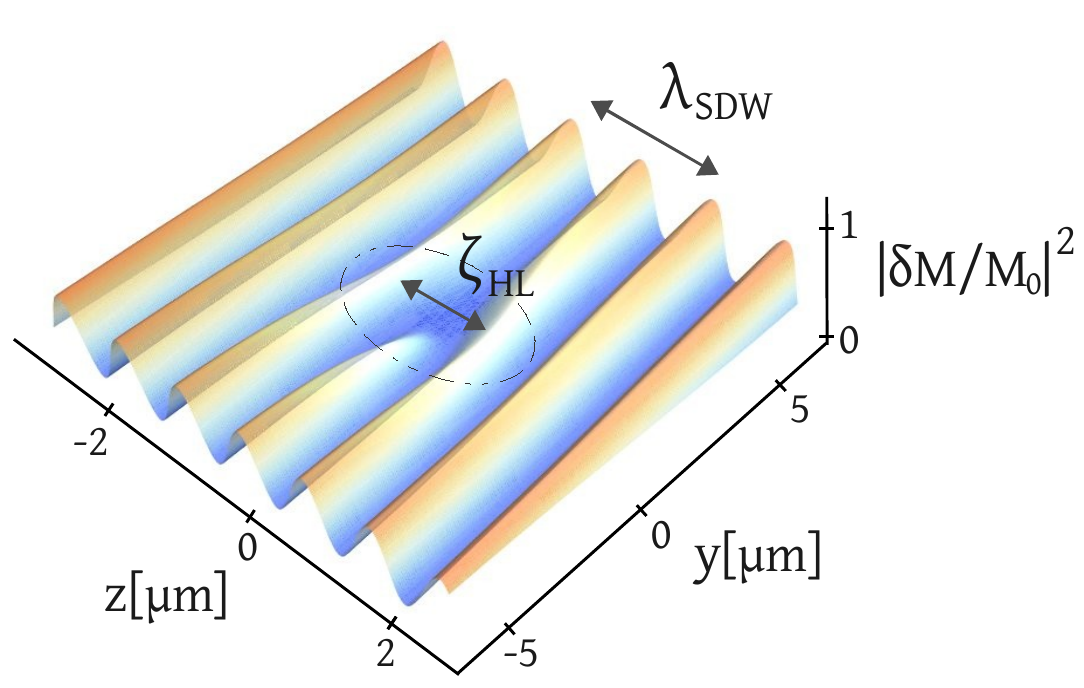} 
   \caption{Vortex structure with an elliptic cross section of aspect ratio $\gamma=\sqrt{m_{||}/m_{\perp}}\sim 5$. Such structure emerge as a dislocation over the spin density wave, and with a Burgers vector proportional to the winding number of the vortex. The presented figure corresponds to the square of magnetization deviation from saturation $|\delta {\vec M}/M_0|^2$ normalized by the magnetization saturation $M_0$.} 
   \label{fig: vortex1}
\end{figure}

In the experiment in Ref. \cite{Demokritov0} the critical chemical potential necessary to achieve condensation is $\mu_C=1.5\times 10^{-2}$[meV]. Here we take the critical pumping power (in units of $eV$) as ${\cal P}_c=\mu_C$, the coefficient $\nu=0.1\mu_C$ (the equivalent of $\nu$ in Eq. ($1$) of \cite{Demokritov0} is denoted by $J$) and the magnon effective mass $m_{||}\approx50m_{\perp}\approx 5\times 10^{-3}m_{He}$. We estimate for the healing length $\zeta_{HL}\approx 0.5[\mu m]$, i.e., the size of the vortex is about four times smaller than the wavelength of the condensed magnons. This value is in agreement with the measurements and the model proposed in \cite{Demokritov0}.

 For the values $\lambda_{SDW}$ and $\zeta_{HL}$ obtained above we solve Eq. (\ref{eq: gross-pitaev}) for a vortex described by $\Phi_{(1,2)}({\bf x})$. The vortex-solution is shown in the Fig. \ref{fig: vortex1} where clearly it can be seen as a dislocation in the SDW pattern. Note that this solution corresponds to a pair of vortices with opposite topological charges and located in the same position. Each vortex exists in the $\pm{\bf k}^{0}-$component, respectively. Due to the different longitudinal and transverse masses, the vortex is anisotropic, with an elliptic cross section of aspect ratio $\gamma=\sqrt{m_{||}/m_{\perp}}\sim 5$, for in-plane magnetic field $\sim 0.1$[T], see Fig. \ref{fig: vortex1}. 

It is worth emphasizing that in order to simulate the experiment, the authors in \cite{Demokritov0} were forced to include a term in the Ginzburg-Landau equations that makes it equivalent with the features of Eq. (\ref{eq: gross-pitaev}). Both equations are peculiar due to the anomalous term proportional to $|\Psi^{*}\rangle$, whose coefficient is denoted $J$ in \cite{Demokritov0} and $\nu$ in Eq. (\ref{eq: gross-pitaev}). 

\section{Internal Josephson Effect}
The notion that a condensate formed by two components (such the valley in our present case) might display Josephson oscillations can be traced back to work in cold atom gases \cite{Hall}. As a first step, we study the so-called internal Josephson effect in $\vec{k}$-space. If we separate the phase difference, $\phi$, between the valleys from  Eq. (\ref{eq: gross-pitaev}), doing 
\beq\label{eq: Internal Josephson Ansatz}
|\Psi\rangle= 
\left(
  \begin{array}{c}
      {\rm e}^{i\phi(t)}\sqrt{n_1(t)} \\
      {\rm e}^{-i\phi(t)}\sqrt{n_2(t)} \\
   \end{array}
\right)
   \eeq  
 where $n_{(1,2)}(t)$ are the density in each valley, we can easily find a Josephson-like relationship. 
Defining the imbalance of magnons density between valleys, 
\beq 
p = \frac{\langle\Psi|\sigma_z|\Psi\rangle}{\langle\Psi|\Psi\rangle}\equiv \frac{n_1-n_2}{n_1+n_2},
\eeq
 the phase difference displays a damped behavior due to the dissipation coefficient $\alpha$. This coefficient couples the equations for $\phi$ and $p$ 
\beqa
\hbar\dot{\phi}&=&-\left(\mu+(\gamma_1+\gamma_2)n\right)p-\hbar \alpha\dot{p},\label{eq: Joseph}\\
\dot{p}&=&-\alpha\dot{\phi} \nonumber
\eeqa
that can be derived from  Eq. (\ref{eq: gross-pitaev}). In this case, Eqs. (\ref{eq: Joseph}) has an interesting implication concerning the damping of the sliding modes. The equations can be rearranged in the form $\dot{p}=-p/\tau_p$, where $\tau_p=\hbar/2\alpha\left(\mu+(\gamma_1+\gamma_2)n\right),$ and correspond to a simple dependence of the damping rate of such modes on the net magnon density of the system. There is a profound relationship between Eq. (\ref{eq: Joseph}), and
the semiclassical equations of motion for the collective dynamics of a Josephson junction \cite{Makhlin} and a single-domain easy-plane ferromagnet in an in-plane field \cite{Rossi2}. This phase relationship  over the relative phase of the condensate indicates that Josephson-related effects should be displayed by the system. Those effects will link the dynamics of magnon population and the spatial configuration of the magnetic patterns.\\
\begin{figure}[htbp] %  figure placement: here, top, bottom, or page
    \centering
    \includegraphics[width=0.9\textwidth]{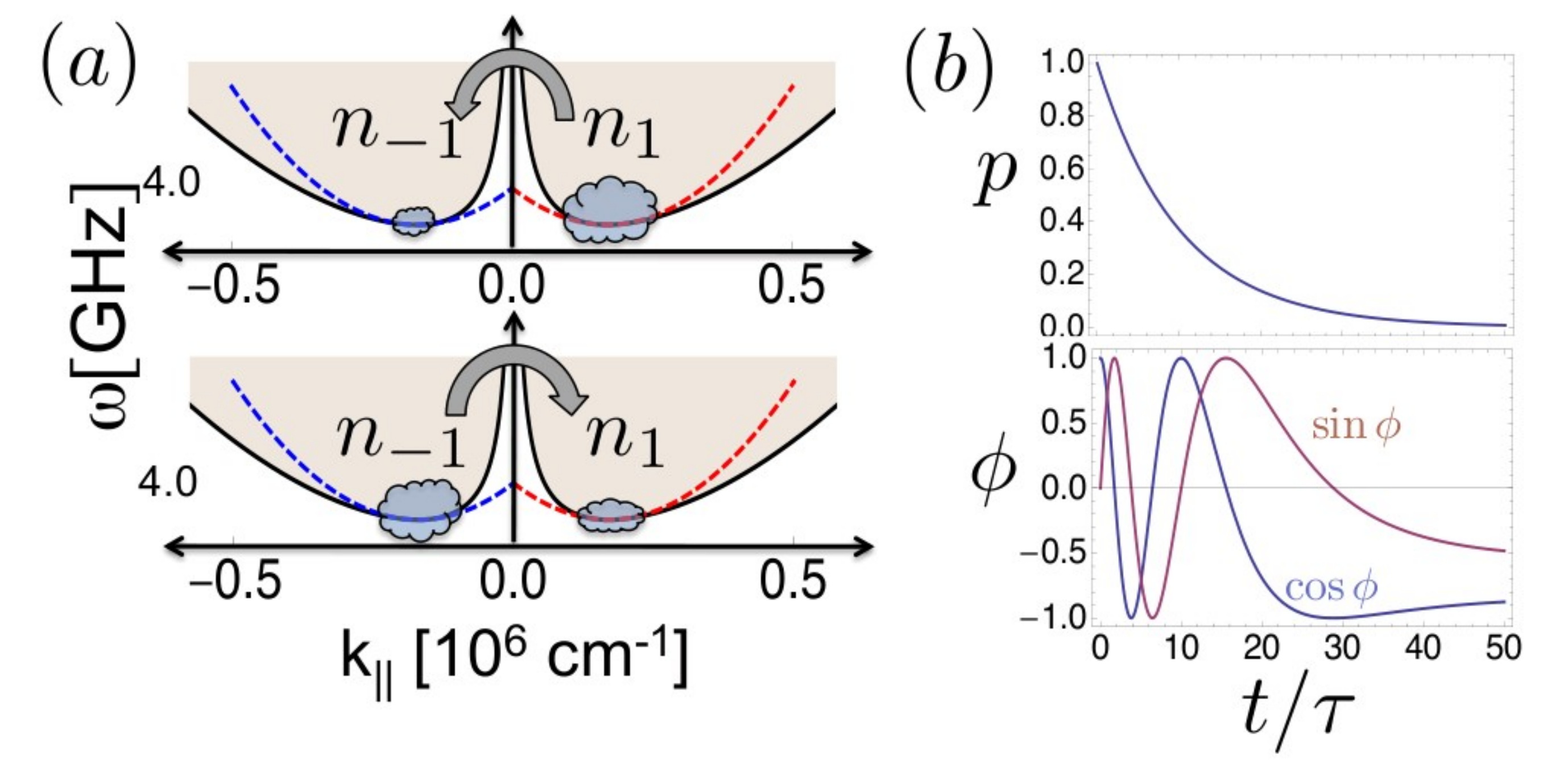} 
    \caption{(a) Illustration of internal Josephson oscillations between valley $\pm {\bf k}_0$, the unbalance between the valleys drives phase oscillations. Damping works restoring the balance between valleys. (b) Numerical solutions of Eq. (\ref{eq: Joseph}) for $\alpha=10^{-4}$, with initial conditions $p(0)=1$ (complete polarization) and $\phi(0)=0$. Such mode characterized by the imbalance population, $p(t)$, and the difference phase $\phi(t)$ between both components of the condensed cloud.}
    \label{fig: sliding}
 \end{figure}

\section{Magnonic Josephson Junction}\label{sec: basic model}
We propose a magnonic Josephson junction (MJJ) on ferromagnetic YIG thin films for weakly linked magnon condensates. The splitting of the cloud condensate can be implemented by applying a direct current, where the tunneling between both states is adjusted by varying the current and geometric parameters of the setup. This approach naturally arises from the experimental studies of spin wave tunneling in a nonuniform magnetized thin films \cite{Demokritov7,Hillebrands1}. In this section we take out the realization of MJJ and will build, from a phenomenological point of view, the semiclassical equations of motion for the collective dynamics between the condensate states. 
 \begin{figure}[htbp] %  figure placement: here, top, bottom, or page
    \centering
   \includegraphics[width=3in]{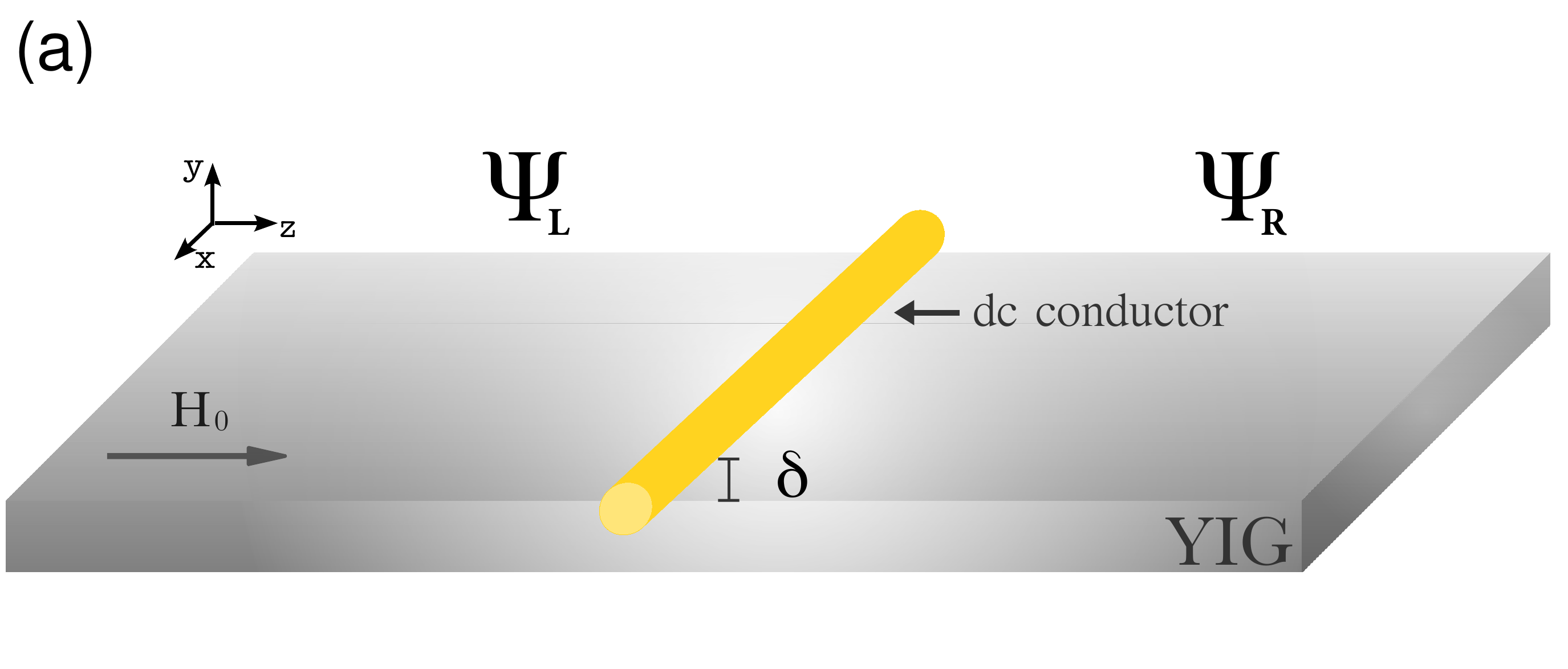}
     \includegraphics[width=2in]{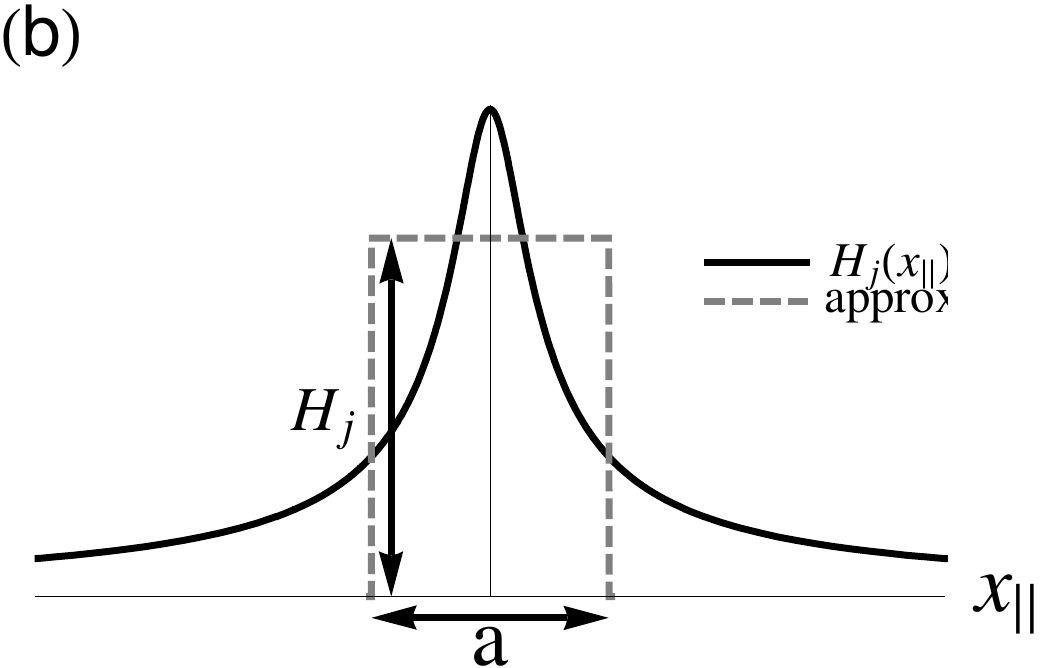}
   \caption{(a) Sketch of the experimental setup for the magnon Josephson's effect realization with a cartoon of the two-mode approximation over the full macroscopic wave function of the condensate state. The spatial fragmentation of the cloud magnon condensate over the YIG thin film is allowed by means of a DC current crossing a wire conductor. (b) The DC current produces a local inhomogeneity, H$_{\textrm{j}}(x_{||})$, in the magnetization. The approximation used to solve Eq. (\ref{eq: gross-pitaev1}) is indicated by the dashed line.}
   \label{fig: experimentalsetup}
\end{figure}

Once the population of magnons, created through a parametric pumping in a YIG thin film, surpasses a critical level given by Eq. (\ref{eq: transition}), the system develops a condensed phase characterized by phase coherence.  When this macroscopic state is partitioned in two condensate clouds, the low-energy dynamics between them is completely described by two macroscopic observables, namely, population imbalance $\eta$ and the relative phase $\phi$. To separate the condensate, we introduce a local inhomogeneity in the magnetization, produced by means of a current that goes through a conductor and placed transversely to the YIG sample. In Fig. \ref{fig: experimentalsetup} a cartoon for the proposed experimental setup is depicted.

As shown above, the condensate state of magnons has two components that belong to the vicinity of energy minimum due to the double valley degeneracy, $\pm{\bf k}^{0}$ in the spectrum Fig. \ref{fig:spectrum}.  The dynamics follows the pseudo-spin GP equation,
\beqa\label{eq: gross-pitaev1}
i\hbar (1+i\alpha)\partial_t|\Psi\rangle&=&-\frac{\hbar^2}{2 m_{||}} \nabla_{||}^2 |\Psi\rangle-\frac{\hbar^2}{2 m_{\perp}} \nabla_{\perp}^2 |\Psi\rangle + \mu |\Psi\rangle+ V_{{\textrm{j}}}({\bf r})|\Psi\rangle + \mbf{\nu}\sigma_{x}|\Psi^*\rangle\nonumber\\&+&\gamma_1 |\Psi|^2|\Psi\rangle+\gamma_2 \langle\Psi|\sigma_z|\Psi\rangle\sigma_z|\Psi\rangle
\eeqa
where the potential barrier produced by the current crossing the wire conductor has the simple form   
$V_{\textrm{j}}(x_{||})=\hat{\gamma}\hbar H_{\textrm{j}}(x_{||})$, with $H_{\textrm{j}}(x_{||})$ the Oersted magnetic field produced by the dc current. For the proposed geometry, Fig. \ref{fig: experimentalsetup}(a), it is given by $2\pi H_{\textrm{j}}(x_{||})={j}/{\sqrt{\delta^2+x_{||}{}^2}}$, where $\delta$ is the separation between the wire and the YIG film and $\hat{\gamma}$ is the effective coupling between magnons and the magnetic field. This external potential introduces an additional energetic gap that the condensed magnons must overcome to get at each side of the barrier, i.e., this plays the role of a weak link between the magnon condensates. To understand the phenomena, predicted by Eqs. (\ref{eq: gross-pitaev1}) we implement the so-called two-mode approximation \cite{Giovanazzi}, writing the full macroscopic wave function as the addition of a two spatially separated time-dependent states, 
\beq\label{eq: twomodeapprox}
\Psi({\bf x},t)=\psi_{\textrm{L}}(t)\Phi_{\textrm{L}}({\bf x})+\psi_{\textrm{R}}(t)\Phi_{\textrm{R}}({\bf x}),
\eeq
where the left and right modes can be obtained from $\Phi_{\textrm{L,R}}({\bf x})=\frac{1}{\sqrt{2}}\left(\Phi_g\pm\Phi_x\right)$, corresponding to the symmetric and antisymmetric functions, which are constructed from the ground-state $\Phi_g$ and the first excited state $\Phi_x$, satisfying the stationary Gross-Pitaevskii equation. This approximation in the Gross-Pitaevskii equation has proven to be a successful description to predict the existence of Josephson tunneling phenomena in clouds of bosonic system confined in a double-well potential \cite{raghavan}. 

Uniformity in the direction parallel to the wire allows a further simplification, the wavefunction depends just on the longitudinal coordinate and the  stationary Gross-Pitaevskii equation can be reduced to the one-dimensional nonlinear Schrodinger equation with an external potential. To determine $\Phi_g$ and $\Phi_x$, we approximate the barrier shape into a piecewise constant potential, as described in Fig. \ref{fig: experimentalsetup}(b), with a strength $H_j$ and a spatial size $a$. We solve the non-linear equations in terms of Jacobi functions \cite{raghavan} considering outside and inside the barrier the elliptic functions $\sn$ and $\nc$ were used, respectively. After matching appropriate boundary conditions, we determine $\Phi_g$ and $\Phi_x$, i.e., we impose that $\Phi_g(x_{||}=\pm a/2)=\Phi_x(x_{||}=\pm a/2)$ and $\Phi'_g(x_{||}=\pm a/2)=\Phi'_x(x_{||}=\pm a/2)$. Evaluating the Eq. (\ref{eq: twomodeapprox}) on the Gross-Pitaevskii equation, Eq. (\ref{eq: gross-pitaev1}), we find the Josephson equations for the two dynamical modes that obey
\beqa\label{eq: josephson1a}
i\hbar\left(1+i\alpha\right)\partial_t\psi_{\textrm{L}}(t)&=&\left[{\rm E}_{\textrm{L}}+{\rm U}_{\textrm{L}}\left(\gamma_1\left|\psi_{\textrm{L}}\right|^2+\gamma_2\left(\psi^{*}_{\textrm{L}}\sigma_z\psi_{\textrm{L}}\right)\sigma_z\right)\right]\psi_{\textrm{L}}(t)\nonumber\\&+&\mbf{\nu}\sigma_x\psi^{*}_{\textrm{L}}(t)+{\rm K}\psi_{\textrm{R}}(t)\\
i\hbar\left(1+i\alpha\right)\partial_t\psi_{\textrm{R}}(t)&=&\left[{\rm E}_{\textrm{R}}+{\rm U}_{\textrm{R}}\left(\gamma_1\left|\psi_{\textrm{R}}\right|^2+\gamma_2\left(\psi^{*}_{\textrm{R}}\sigma_z\psi_{\textrm{R}}\right)\sigma_z\right)\right]\psi_{\textrm{R}}(t)\nonumber\\&+&\mbf{\nu}\sigma_x\psi^{*}_{\textrm{R}}(t)+{\rm K}\psi_{\textrm{L}}(t),\nonumber
\eeqa
where the spatial dependence was integrated utilizing the orthogonality condition for $\Phi_{\textrm{L,R}}({\bf x})$. This system of nonlinear equations represents the dynamics between two magnon condensate states with a coupling factor, proportional to the wave function overlap. The information about of the spatial dependence is contained in the coefficients ${\rm E}_{i}$, ${\rm U}_{i}$ and ${\rm K}$. The meaning of such parameters is the following: the coefficient ${\rm E}_{i}$ represents the zero point energy in each region, ${\rm U}_{i}n^{\pm{\bf k}^{0}}_{i}$ are proportional to the self-interaction energies, while ${\rm K}$ describes the amplitude of the tunneling between both condensates. Those coefficients are written in terms of $\Phi_{g,x}$ and the effective parameters that characterize the condensed phase, whose expressions for each one correspond to
\beqa\label{eq: josephsoncoefficients}
{\rm E}_{i}&=&\int d{\bf r}\Phi_{i}({\bf r})\left[-\frac{\hbar^2}{2m}\nabla^2+\mu+V_{\textrm{j}}({\bf r})\right]\Phi_{i}({\bf r})\\
{\rm U}_{i}&=&\int d{\bf r} \left|\Phi_{i}(\bf r)\right|^4\\
{\rm K}&=&\int d{\bf r}\left(\frac{\hbar^2}{2m}\nabla\Phi_{\textrm{L}}({\bf r})\nabla\Phi_{\textrm{R}}({\bf r})+V_{\textrm{j}}({\bf r})\Phi_{\textrm{L}}({\bf r})\Phi_{\textrm{R}}({\bf r})\right)
\eeqa
where the weakly linked approximation was used. Clearly the Josephson coefficients are defined by the geometry of the barrier and the microscopic parameters of the condensed phase.

\section{Phase dynamics in a Magnonic Josephson Junction}
In the previous section we have stablished several properties of the magnon condensate fragmented by a potential well. To study the dynamics between the condensates, hereinafter we will parameterize the wave functions $\psi_{i}$, $i=\textrm{L},{\textrm{R}}$ in terms of the occupation density and its phase as; 
\beq\label{eq: twomodewavefunction1}
\psi_i=\left(\begin{array}{c}\sqrt{n_i(t)}e^{i\phi_i(t)}\\\sqrt{n_i(t)}e^{-i\phi_i(t)}\end{array}\right).
\eeq
Note that we will restrict ourselves to the case where the internal oscillations are frozen, i.e., the only variables of interest are the oscillations between both left and right states. We call this behavior the external magnon Josephson effect. The interesting case on the dynamics of the interplay between the $\pm{\bf k}^{0}$-valleys and the spatially separated clouds is left for a further study. 

Writing the two-mode dynamical equation, Eq. (\ref{eq: josephson1a}), using the occupation density-phase representation for the wave functions, we obtain that the coupling among magnon condensates are described by: 
\beqa\label{eq: magnonjosephsonjunction}
\dot{\eta}&=&-\alpha\Gamma\eta-\sqrt{1-\eta^2}\sin\phi\\
\dot{\phi}&=&\Lambda\eta+\frac{1}{\sqrt{1-\eta^2}}\left(\eta\cos\phi-\alpha\sin\phi\right) \nonumber
\eeqa
where the magnon population imbalance and relative phase are defined as $\phi\equiv \phi_{\textrm{R}}(t)-\phi_{\textrm{L}}(t)$ and $\eta\equiv\left(n_{\textrm{L}}(t)-n_{\textrm{R}}(t)\right)/n_{\textrm{T}}$. The time is rescaled to a dimensionless characteristic time $t_c=\hbar/2{\rm K}$, which is  typically $\sim 10$ $[$ns$]$ for the values of the coefficients used in this article, and the phenomenological parameters enter in the following combinations
\beqa
\Lambda&=&\gamma_1{\rm U}\rho_c/2{\rm K},\nonumber\\
\Gamma&=&\left(2{\rm E}+2\gamma_1{\rm U}\rho_c+\bar{\nu}-\mu\right)/2{\rm K}.\nonumber
\eeqa

The pair of equations Eq. (\ref{eq: magnonjosephsonjunction}) represents the macroscopic interference  between the clouds of magnons. In the dissipationless limit, the full dynamics is determined exclusively by the parameter $\Lambda$ and, as was pointed in Ref. \cite{Giovanazzi}, that Eq. (\ref{eq: magnonjosephsonjunction}) can be derived as the hamiltonian equations associated to the total conserved energy ${\cal H}[\phi,\eta]={\Lambda}\eta^2/2-\sqrt{1-\eta^2}\cos\phi$.
\begin{figure}[htbp] 
   \centering
   \includegraphics[width=0.9\textwidth]{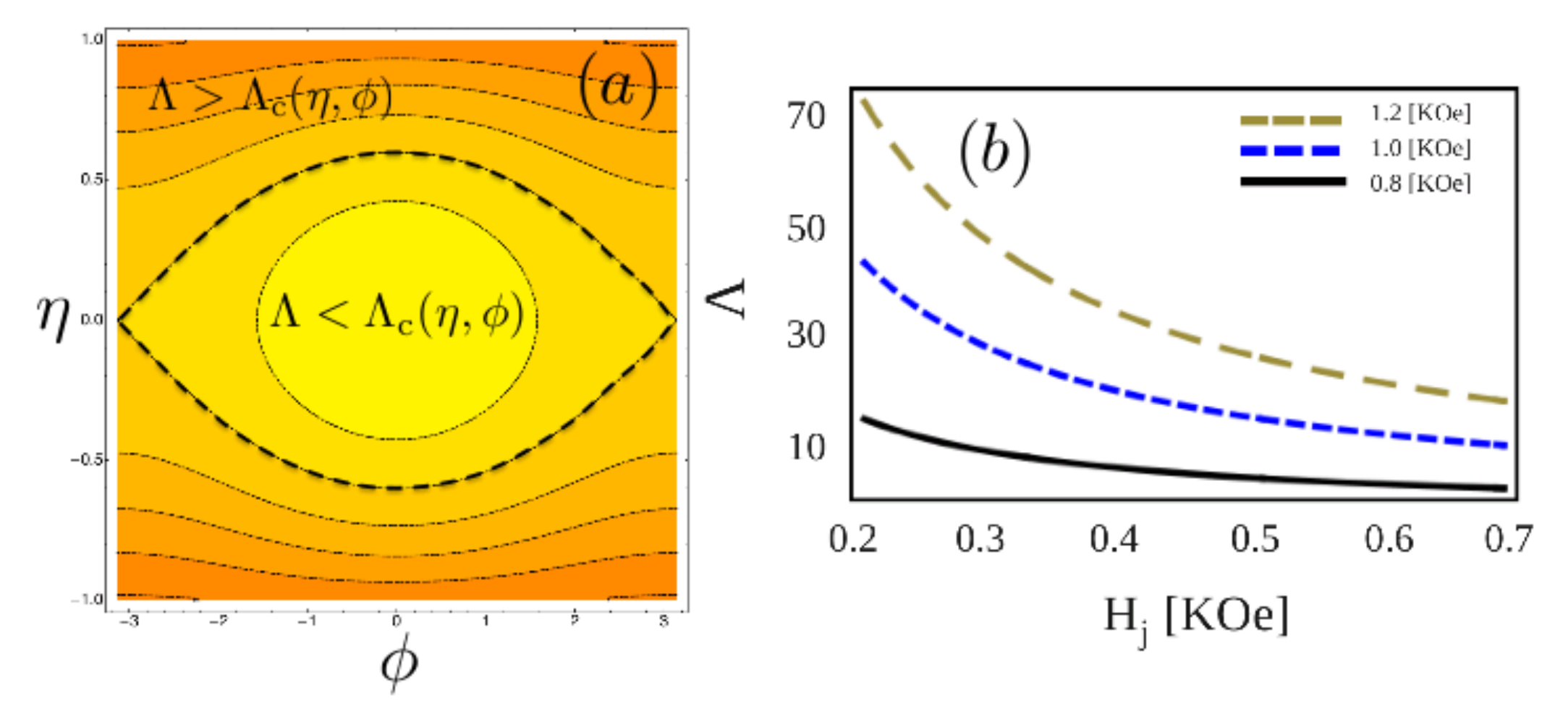}
   \caption{(a) Energy levels in phase space for the magnonic Josephson oscillations. The close or open nature of the iso-energy curves delimits two, qualitative different, Josephson's oscillation regimes.  The pictures represents the case $\Lambda\sim 10$. (b) Parameter $\Lambda$ calculated from our model for the magnonic junction as a function of the  Oersted field generated by the current.}
   \label{fig: phasediagram}
\end{figure}

%As an example of the dynamics encoded in Eq. (\ref{eq: magnonjosephsonjunction}) we start with the most elementary case, i.e. the Josephson's oscillations of small amplitude, whose behavior reveals the macroscopic interference due to the interaction of the condensates of magnons. 
In Fig. \ref{fig: acMagnonJosephson1}(a) and \ref{fig: acMagnonJosephson1}(b) we illustrate the solutions for the magnon current and relative phase considering small and large amplitude oscillations, respectively. The solutions are shown for a dissipation $\alpha=10^{-4}$ \cite{Kajiwara}, an in-plane magnetic field $H_{0}=1[\textrm{KOe}]$ \cite{Demokritov2} and considering values for the inhomogeneity in the magnetization in the range $250$[Oe]$<H_{\textrm{j}}<500$[Oe]. In the small amplitude limit the frequency of oscillations satisfy the simple form $\omega_{\textrm{ac}}=\sqrt{1+\Lambda}$. In Fig. \ref{fig: acMagnonJosephson1}(a) we plot the small amplitude oscillations for $\Lambda=30$($H_{\textrm{j}}=300[\textrm{Oe}]$) and initial conditions $\eta_0=0.1$, $\phi_0=0.1$. The case of large amplitude oscillations, Fig. \ref{fig: acMagnonJosephson1}(b),  is plotted using $\Lambda=30$($H_{\textrm{j}}=300[\textrm{Oe}]$) and initial conditions, $\eta_0=0.1$, $\phi_0=0.8\pi$. A feature that distinguishes this regime of oscillations is the vanishing mean value of the magnon current $\langle\eta(t)\rangle=0$, i.e., the system of two macroscopic states oscillates around of the equilibrium value set in $\eta_{\textrm{eq}}=0$. Moreover, it follows from Fig. \ref{fig: acMagnonJosephson1}(b) that the large-amplitude oscillations are quickly damped, respect to the small amplitude regime.

The solutions presented here correspond to the Josephson's effect for the magnon condensate. The frequency of oscillations of the magnon current is directly related to the macroscopic relative phase of the condensates, which are within the typical experimental resolution range \cite{BLS}. This provides a precise methodology for the observation of ac Josephson's oscillations over the YIG thin films and clarifying the true nature of the magnon cloud.

\section{Macroscopic self-trapping of magnons}
The magnon Josephson's oscillations discussed so far correspond to an oscillating current of magnons crossing the potential barrier and characterized by symmetric  oscillations of the occupation density $\langle\eta(t)\rangle=0$. However, this scenery changes drastically when the interaction surpasses a critical value $\Lambda>\Lambda_c$, where the Josephon's oscillations follow a qualitatively different behavior. In this regime the evolution of the magnon population imbalance is characterized by a nonzero time-average $\langle\eta(t)\rangle\neq0$, i.e., there is a self-trapping of magnons at one side of the wire. Furthermore, In this regime the dynamics is strongly influenced by the initial conditions.
\begin{figure}[htbp] 
   \centering
   \includegraphics[width=1.\textwidth]{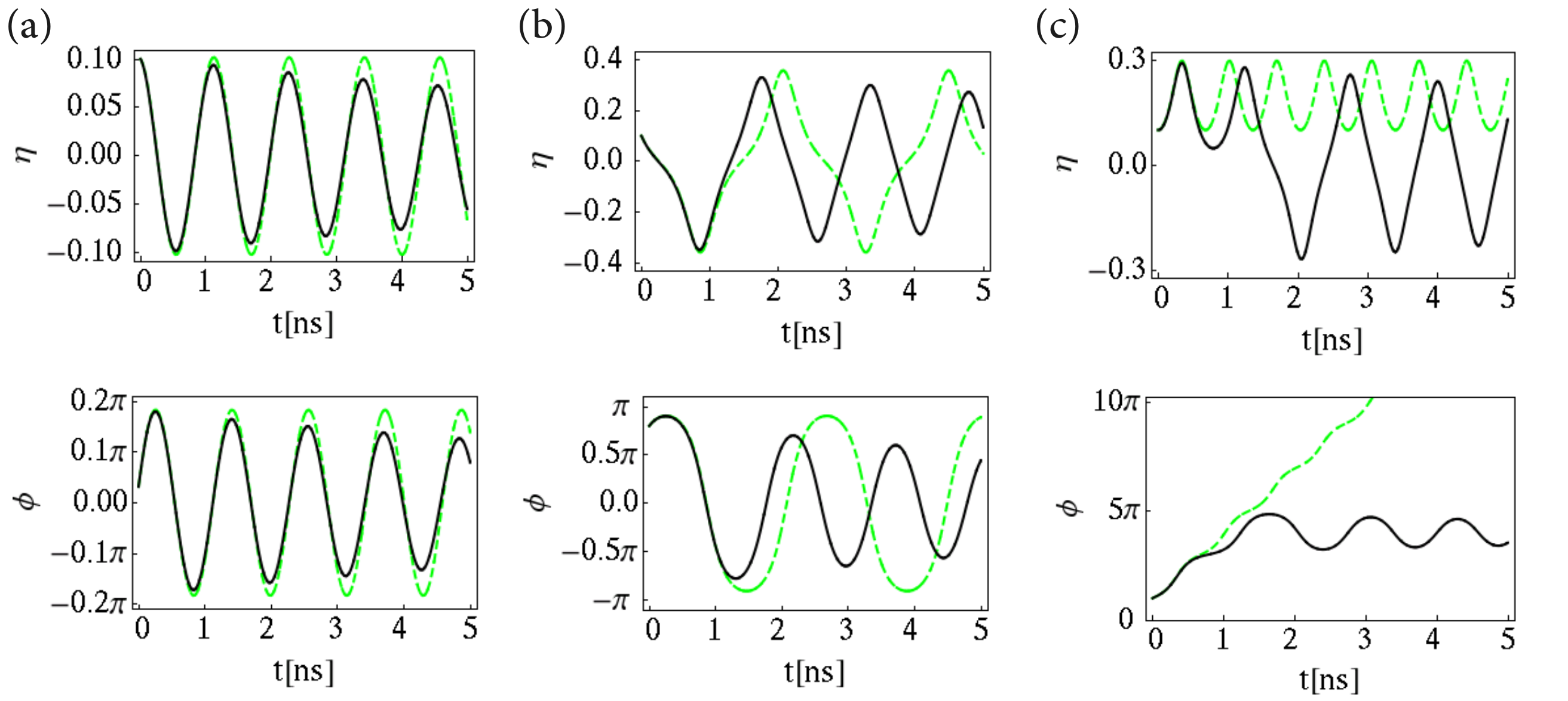}
   \caption{Dynamical behavior of both relative phase $\phi$ and population imbalance $\eta$ between the clouds magnons condensate. The solutions are calculated for the typical experimental conditions over the \textrm{YIG} thin film, an in-plane magnetic field $H_0=1$[KOe] and dissipation $\alpha=10^{-4}$. In (a) and (b)  the small amplitude and long wave oscillations are displayed for $\Lambda(H_j=300[\textrm{Oe}])=30$ and initial conditions, $\phi(0)=0.1\pi$-$\eta(0)=0.1$ and $\phi(0)=0.8\pi$-$\eta(0)=0.1$, respectively. While in (c) the MST solutions are obtained for $\Lambda(H_j=200[\textrm{Oe}])=50$ and initial conditions $\phi(0)=\pi$, $\eta(0)=0.1$.}
   \label{fig: acMagnonJosephson1}
\end{figure}

This nonlinear phenomenon was discovered in the context of BEC's of alkali gases \cite{raghavan, Levy} and is coined as macroscopic quantum self-trapping (MQST), whose quantum nature involves the coherence of a macroscopic number of bosons in the pair of condensates. 
%Here we show that the condensate of magnons manifest themselves in a macroscopic self-trapping state (MST), for certain values of the self-interaction and where the dissipation play an important role in the dynamics of such state. 
In our problem that regime is achieved when the parameter $\Lambda$ exceeds a critical value $\Lambda_c$, which is in turn specified by the initial conditions $\eta_0$ and $\phi_0$, through: 
 \beq
\Lambda_c=\frac{1+\sqrt{1-\eta_0^2}\cos\phi_0}{\eta_0^2/2}.
\eeq
Once $\Lambda>\Lambda_c$ is satisfied, the Josephson's oscillations are driven to the MQST regime. In Fig. \ref{fig: acMagnonJosephson1}(c) is shown this solution for $\Lambda(H_j=200[\textrm{Oe}])=50$ and initial conditions $\phi(0)=\pi$, $\eta(0)=0.1$. Due to the effect of damping, it can be seen that the condensed magnons remain in the MQST state for some time until decay into a long-amplitude oscillation. Indeed the dissipation quickly destroys such state giving way to the Josephson's oscillations studied before. That point can be visualized in the breakdown of the MQST state, establishing a characteristic life-time for this state.

\section{Discussion and summary}\label{sec: conclusions}
Before closing it is worth commenting on the experimental signals that might be expected from the effects discussed in this work. Magnetization oscillations can be measured in several ways. The basic mechanism used so far in the context of magnon condensates is the Brillouin light scattering technique (BLS) \cite{BLS}. Such technique  probes the magnons system by studying their effect on microwave radiation reflected by the sample. In this way it might be expected that the oscillations in magnon density between the two magnonic clouds might be detected. Our predictions involve oscillation periods of the order of $5-20$[ns]. Such oscillations are however shorter than the characteristic resolution of the BLS measurement. As an alternative the magnon dynamics can be transformed into spin currents pumped into a metallic sample in contact with the system \cite{Tserkovnyak}. Such currents have been measured by means of the inverse spin Hall effect in Pt \cite{ISHE} that converts them in charge currents. In the present case, it is easy to show that the presence of  magnon-condensate implies a constant current. Oscillations in such current can be detected and interpreted as signals of the underlying oscillations.

In conclusion we have presented a phenomenological theory that, focusing only on the low-energy and momentum projections of the magnon spectrum, accounts for the collective dynamics of a Bose-Einstein condensate of magnons. Such theory has allowed us to both provide a simple understanding of the mechanisms behind the condensation of magnons and to establish a clear understanding of the meaning of the collective wave function used to describe it. 
%Despite these efforts to understand the BEC phase in YIG a systematic study concerning the interaction between the condensed and thermal magnons is still lacking.
In terms of such description we discuss the nature of macroscopic interference between magnon clouds. Starting with the discussion of the internal Josephson oscillations, that correspond to oscillations between the $\pm k^{0}$ components of the condensed cloud, we have highlighted the close relation between such effects and the well-known Josephson effect. Using those ideas, we presented a detailed calculation of the Josephson oscillations between two magnon clouds, spatially separated in a magnonic Josephson junction. Among the results, we remark the clear and distinctive oscillations that characterize common Josephson oscillations and also a regime that corresponds to the so-called macroscopic quantum self-trapping, that locks the oscillations favoring one side of the junction over the other. 

Despite of growing efforts to study BEC of magnons, is still exist a controversy about the real nature of the condensate of magnons. It has been claimed in \cite{Ruckriegel} that the condensate of magnons doesn't admit the Onsager and Penrose criterium \cite{Penrose} for the Bose-Einstein condensation. In fact, is argued that the condensate of magnons is merely a reorientation of the macroscopic magnetization and corresponding just to a Ising transition in a magnet, where the Ising transition is associated to the breaking of the $Z_2$-symmetry. However, we think that the condensate of magnons meets all the characteristics of an usual condensate, in agreement with Ref. \cite{Troncoso}. We claim that, besides the $Z_2$ symmetry, a residual $U(1)$ symmetry still exists, and therefore, the pumped magnon gas has the symmetry $Z_2 \times U(1)$. We have pointed out that the breaking of the residual symmetry leads to the Bose-Einstein condensation of magnons. Therefore, the problem of macroscopic tunneling between condensates of magnons is of importance to both explore the superfluid properties of the magnon condensate and to establish a way to settle the controversy from the experimental point of view.

\section{Acknowledgements} ASN would like to thank Professor R. A. Duine for helpful comments. This work was partially funded by Proyecto 
Fondecyt numbers 11070008 and 1110271, Proyecto Basal FB0807-CEDENNA, N\'ucleo Cient\'ifico Milenio P06022-F and Proyecto Anillo de Ciencia y Tecnología, ACT 1117.\\

\bibliographystyle{elsarticle-num}

\end{document}